\documentclass[sigconf]{acmart}
\usepackage{float}
\usepackage{multirow}
\usepackage{multicol}
\usepackage{booktabs} 
\usepackage{caption}
\usepackage{graphicx}
\usepackage{textcomp}
\usepackage{xcolor}
\usepackage{color,soul}

\usepackage{stfloats}
\usepackage{microtype}
\usepackage[labelformat=simple]{subcaption}

\usepackage{hyperref}
\usepackage{csquotes}
\usepackage{caption, subcaption}

\usepackage{breqn}
\usepackage{balance}
\usepackage{tabu}                     
\usepackage{multirow}                 
\usepackage{multicol}                 
\usepackage{multirow}              
\usepackage{float}
\usepackage{makecell}           
\usepackage{booktabs}
\usepackage{enumitem}
\usepackage{graphicx} 
\usepackage{float}
\usepackage{pgfplots} 
\pgfplotsset{width=8cm,compat=1.9}
\usepackage{url}
\usepackage{multirow}
\usepackage{diagbox}
\usepackage{adjustbox}
\usepackage{booktabs}
\usepackage{tabularx}
\usepackage{wrapfig}
\usepackage{float}
\usepackage{amsmath}
\usepackage{tikz}
\usepackage{tcolorbox}
\usepackage{lipsum}
\usepackage{setspace}
\usepackage{stfloats}

\pdfoutput=1

\newcommand{\circled}[1]{%
  \begin{tikzpicture}[baseline=(char.base)]
    \node[shape=circle, draw, inner sep=1pt] (char) {#1};
  \end{tikzpicture}%
}

\usepackage[ruled,vlined]{algorithm2e}
\usepackage{longtable}

\DeclareMathOperator*{\argmin}{arg\,min}

\AtBeginDocument{%
  }

\begin{document}

\title{Mitigating Unauthorized Speech Synthesis for Voice Protection}


\author{Zhisheng Zhang$^{\clubsuit}$, Qianyi Yang$^{\clubsuit}$, Derui Wang$^{\diamondsuit}$, Pengyang Huang$^{\clubsuit}$, 
Yuxin Cao$^{\star}$, Kai Ye$^{\heartsuit}$, Jie Hao$^{\clubsuit}$}
\def 
\authors{Zhisheng Zhang, Qianyi Yang, Derui Wang, Pengyang Huang, Yuxin Cao, Kai Ye, Jie Hao}
\authornote{Jie Hao is the corresponding author.}
\affiliation{%
    \institution{$^{\clubsuit}$Beijing University of Posts and Telecommunications
    \city{Beijing}
    \country{China}
}
}
\affiliation{%
    \institution{$^{\diamondsuit}$CSIRO's Data61
    \city{Melbourne}
    \country{Australia}
}
}
\affiliation{%
    \institution{$^{\star}$National University of Singapore
    \city{Singapore}
    \country{Singapore}
}
}
\affiliation{%
    \institution{$^{\heartsuit}$The University of Hong Kong
    \city{Hong Kong SAR}
    \country{China}
}
}

\email{{zzs2002, youmu314, huangpengyang, haojie}@bupt.edu.cn,}
\email{derek.wang@data61.csiro.au, yuxincao@comp.nus.edu.sg, yek21@hku.hk}

\settopmatter{printacmref=false}
\renewcommand\footnotetextcopyrightpermission[1]{}
\pagestyle{plain}

\begin{abstract}
  With just a few speech samples, it is possible to perfectly replicate a speaker's voice in recent years, while malicious voice exploitation (\textit{e.g.}, telecom fraud for illegal financial gain) has brought huge hazards in our daily lives. Therefore, it is crucial to protect publicly accessible speech data that contains sensitive information, such as personal voiceprints. Most previous defense methods have focused on spoofing speaker verification systems in timbre similarity but the synthesized deepfake speech is still of high quality.
  In response to the rising hazards, we devise an effective, transferable, and robust proactive protection technology named \textbf{P}ivotal \textbf{O}bjective \textbf{P}erturbation (POP) that applies imperceptible error-minimizing noises on original speech samples to prevent them from being effectively learned for text-to-speech (TTS) synthesis models so that high-quality deepfake speeches cannot be generated. 
  We conduct extensive experiments on state-of-the-art (SOTA) TTS models utilizing objective and subjective metrics to comprehensively evaluate our proposed method. The experimental results demonstrate outstanding effectiveness and transferability across various models. Compared to the speech unclarity score of 21.94\% from voice synthesizers trained on samples without protection, POP-protected samples significantly increase it to 127.31\%. Moreover, our method shows robustness against noise reduction and data augmentation techniques, thereby greatly reducing potential hazards.~\footnote{Our code is available at \url{https://github.com/wxzyd123/Pivotal_Objective_Perturbation}.}
\end{abstract}




\maketitle

\begin{figure}[t]
\centerline{\includegraphics[width=0.4\textwidth]{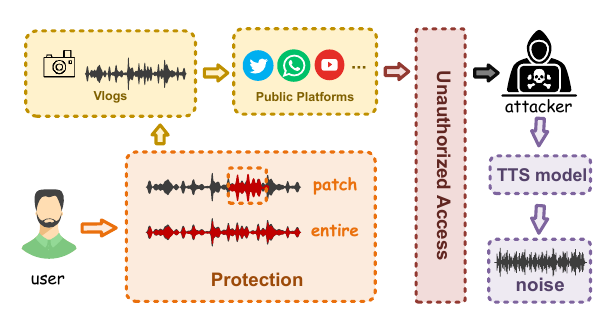}}
\caption{The protection scenario in the real world. The attackers cannot synthesize usable speeches after obtaining protected personal speeches without authorization.}
\label{fig_scenario}
\end{figure}

\section{Introduction}\label{section_intro}
Fueled by advancements in generative artificial intelligence (AI), voice cloning has become a hot topic in recent years. Well-trained speech synthesis models can now synthesize realistic speech with specific text and speaker features (\textit{e.g.}, speaker ID, and some simple speech samples) at test time. These advancements stem from the development of deep neural networks (DNNs), leading to models like Tacotron2~\cite{shen2018natural} and Transformer-TTS~\cite{li2019neural} for single-speaker synthesis, and GlowTTS~\cite{kim2020glow}, VITS~\cite{kim2021conditional} and MB-iSTFT-VITS~\cite{kawamura2023lightweight} for multi-speaker scenarios with awesome synthetic speech. As a result, the synthesized speech produced by these models is becoming increasingly indistinguishable from real human speech, faithfully replicating timbre, speaking rhythm, and pitch. Moreover, some zero-shot TTS models, such as YourTTS~\cite{casanova2022yourtts} and MegaTTS2~\cite{jiang2024megatts}, have been proposed with the capacity to clone a speaker's voice from one short speech sample. Fine-tuning these pre-trained models can achieve superior performance with fewer computing and data resources compared to training from scratch.

\noindent\textbf{Voice Protection.} The synthesized speech referred to as deepfake audio is challenging to differentiate from real speech without proper attention, posing significant data security risks. Moreover, in recent years, large language models (LLMs)~\cite{touvron2023llama, achiam2023gpt} have been continuously developed, and their high-quality human-like text output has led to widespread usage among the public. However, LLMs can generate human-like text responses for TTS models, which further creates a need for voice data protection, as voiceprint becomes increasingly important for distinguishing these semantically enhanced synthetic content. Unauthorized exploitation of public audio files for malicious voice synthesis on the internet by attackers can lead to serious security and legal issues, including social engineering spoofing, illegal authentication, and telecom fraud with lethal consequences.

\noindent{\textbf{Existing Defense Methods.}}  
To deal with the hazards of deepfake audio, the existing defensive methods mainly focus on detection and prevention techniques. Detection technologies are usually aimed at cases where the deepfake audio has been synthesized. Some auxiliary measures are adopted to determine whether the audio is generated, such as using MFCC features~\cite{hamza2022deepfake} and other acoustic signal analysis~\cite{blue2022you} simulating auditory effects of the human ear, or by supplementary methods like emotion recognition~\cite{conti2022deepfake} and speaker verification~\cite{pianese2022deepfake}. Detecting the speech~\cite{liu2023protecting} can improve people's awareness of preventing deepfake audio, but the reactive nature of detection usually implies that the audio has already been exploited and cannot achieve protection at the data level. 
In recent years, researchers have proposed methods~\cite{yu2023antifake, huang2021defending, wang2023vsmask} to protect voice at the data level so that similar deepfake speeches cannot be generated and Figure \ref{fig_scenario} shows the scenario of voice anti-cloning. They employ adversarial perturbation to protect the original audio, thereby preventing the model from cloning during the inference procedure.
However, the prevention-based approach achieves speech dissimilarity to spoof speaker verification systems, but the synthetic speeches remain highly usable for test-time protection.
These solutions failed to address the root cause of the deepfake audio problem which is the illegal exploitation of publicly available audio samples.
In contrast, we aim to radically mitigate the threat by making our voice data unlearnable for TTS models. As a result, TTS models trained on the protected data can only produce low-quality speech to prevent a flood search for victims.

\noindent\textbf{Challenges.} To mitigate this deepfake problem in the TTS field, we aim to devise an effective and transferable data protection method so that the protected data cannot be learned from TTS models by making a slight change to the original data. However, compared to the defense methods in other scenarios, there are three challenges specifically in speech synthesis: (1) The inputs of the model are multimodal, where the inputs include text, audio, and spectrograms; (2) Furthermore, the structure of the model is more complex and the defense mechanism should be transferable enough across various models; (3) There are weighted objective functions for optimization, such as duration time, timbre, and style. If the model's objective functions are crafted for perturbation generation, the performance cannot be better because perturbation can not affect unrelated objective functions.

\noindent\textbf{Our Defense Strategy.} In response to the challenges in the TTS field, we propose a perturbative data protection strategy named Pivotal Objective Perturbation which embeds an imperceptible error-minimizing noise on the original waveform while preserving textual content to fool TTS models. In the design of POP, we analyze a common feature that generative TTS models take audio (or spectrogram) as a part of the outputs and emphasize the main component of a multi-objective function. Based on this common property, we ensure the high transferability of the POP method. Moreover, we consider a fixed-position perturbation~\cite{gokul2024poscuda} for a small patch in POP, which has faster noise generation and better imperceptibility compared to the whole segment perturbation. To verify the effectiveness of our methods, we conduct experiments on multiple datasets and the advanced TTS models and evaluate the protection performance by subjective and objective metrics. The extensive experimental results show high protection effectiveness and transferability of our protected method across models. Compared to the unclarity score of 21.94\% from voice synthesizers trained on speech samples without protection, POP-protected samples significantly increase the speech unclarity to 127.31\% reflecting the great unusability. Furthermore, the POP-protected dataset can defend the perturbation removal and speech augmentation techniques, which show high robustness and guarantee real-world application.

In summary, our paper makes the following contributions:

\begin{itemize} [topsep=0pt]
\item We devise a speech protection method POP that embeds imperceptible perturbation on original audio rendering the protected audio cannot be learned for TTS models against unauthorized and malicious speech synthesis.
\item We analyze the reasons for the design of the POP strategy and specify the principles of our objective function selection with an example using a backbone TTS model.
\item We conduct extensive experiments on current advanced TTS models and datasets. The results demonstrate the great effectiveness, transferability, and robustness of the protected audio on both objective and subjective metrics.
\end{itemize} 

\section{Related Work}\label{section_related}
\subsection{Text-To-Speech Synthesis}   

Modern text-to-speech systems use the basic framework of statistical parametric speech synthesis (SPSS), which consists of three main core components: a text analysis module, an acoustic model, and a vocoder~\cite{zhang2021survey, black2007statistical}.

The text analysis module is responsible for transforming the input text into linguistic features suitable for further processing. Traditional SPSS typically employs Hidden Markov Models (HMM) for acoustic modelling~\cite{yoshimura1999simultaneous, zen2009statistical}, while modern TTS systems moving to more efficient neural network models instead.\cite{zen2013statistical, qian2014training, zen2015unidirectional, fan2014tts}. The vocoder is responsible for synthesizing the generated spectrogram into audio. Early DNN-based synthesis methods tended to perform poorly on long sentences, often dropping words. To overcome these challenges, researchers have introduced alignment mechanisms~\cite{badlani2022one, tachibana2018efficiently}. Among these, autoregressive models achieve alignment through the attention mechanism~\cite{wang2017tacotron, shen2018natural, tachibana2018efficiently}, while non-autoregressive models achieve alignment more efficiently in other ways, such as the Montreal Forced Aligner (MFA)~\cite{mcauliffe2017montreal, ren2021fastspeech, chen2021adaspeech}. The end-to-end TTS system alleviates the problem of one-to-many mapping and further improves speech synthesis~\cite{ren2021fastspeech, shen2018natural}.

\subsection{Privacy Preservation}
The abuse of speech synthesis leaks the privacy of the speaker, and there are some ways to defend against it, such as speaker anonymization and data-based protection.

Speaker anonymization~\cite{fang2019speaker, yoo2020speaker} is an effective way to protect the privacy of the speaker, and the simplest and most direct method is to convert the original speech to text before resynthesizing it through the TTS model to completely remove the voiceprint. In addition, Fang \textit{et al.}~\cite{fang2019speaker} and Han \textit{et al.}~\cite{han2020voice} to hide voiceprint by modifying the x-vector of the original waveform, Yao \textit{et al.}~\cite{yao2024distinctive} can effectively hide the identity of the speaker by modeling the speaker as a matrix and then modifying the values of the matrix.
The existing data-based approach~\cite{huang2021defending, wang2023vsmask, yu2023antifake, li2023voice}, leverages protective algorithms to make some perturbations or changes to the original audio data to achieve voice anti-cloning.

However, both of the current privacy preservation methods have certain limitations, that is, their results bypass the speaker verification systems and achieve dissimilarity in voice quality, but the synthetic audio after model training still has usability. This may lead to attackers obtaining deepfake audio and replacing predetermined targets, thus flooding potential victims. Therefore, we hope to ensure that the deepfake audio is not similar and low-usability.

\begin{figure*}[t]
\centerline{\includegraphics[width=0.85\textwidth]{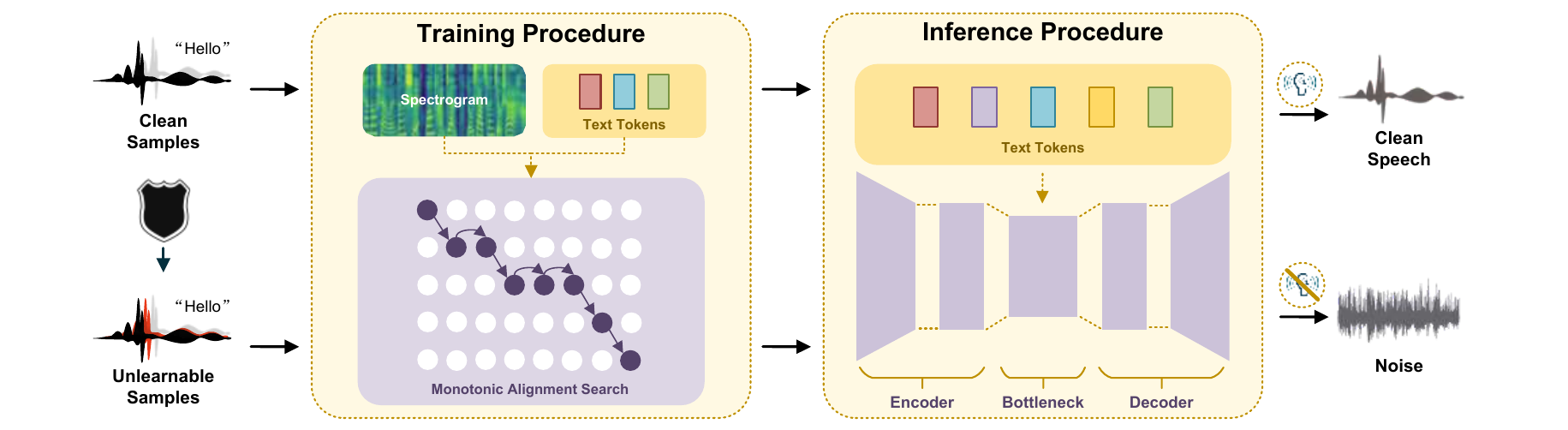}}
\caption{The synthetic results when TTS models trained on clean and protected protected samples respectively.}
\label{fig_workflow}
\end{figure*}

\section{Threat Model}\label{section_threat}
In this section, we outline the capability and limitations of the defender and adversary.

\subsection{Defender Capability and Limitations}. To achieve comprehensive and effective protection of released data at the source stage, our audio data protection scenario is that before users publish their audio files on public platforms such as social media, they can leverage our designed data protection strategy for privacy-preserving. Therefore, the adversary cannot obtain the sensitive unprotected original audio. In this paper, the defender and users perturb the audio at a fixed position for better effect. To more accurately simulate real-world training scenarios, we restrict the defender's knowledge: they are not privy to the specific Text-to-Speech (TTS) model that the adversary might employ for speech synthesis. This requires the defender to fully consider the transferability and robustness of the generated perturbation across different TTS models when designing the protection method, that is, to choose a surrogate model and generate noise that still has high generalization on other models.

\subsection{Adversary Capability.} 
The advanced TTS models can synthesize realistic voices after training clean samples from victims. We consider the two capabilities of the adversary in the real world.

\noindent\textbf{Capability of Data Access.} With the continuous development of internet technology, people are more likely to share their videos publicly on the network, such as vlogs. These videos containing sensitive information can usually be directly downloaded, and adversaries can obtain unauthorized data through techniques (\textit{e.g.}, web crawler), from social media platforms such as YouTube, TikTok, and Facebook.

\noindent\textbf{Capability of Speech Synthesis.} When the adversaries obtain our protected audio, they can train different TTS models with conventional settings and they can use various models for effective speech synthesis training. At the same time, due to the lack of \textit{a prior} knowledge about the obtained audio, they may obverse the embed perturbation after protection. Therefore, the adversary may employ some perturbation removal and data transformation techniques, to disrupt the structural information or backdoor information of the added perturbation, to achieve high-quality speech synthesis.

\section{Overview of Methods}\label{section_methods}
Figure \ref{fig_workflow} illustrates our aim of voice anti-cloning and the protective result of our methods.
In this section, we illustrate the problem definition of unlearnable audio and introduce the primary methods including the speaker selection strategy for fine-tuning and the specific perturbation generation method for TTS systems.

\subsection{Problem Definition}\label{section_definition}
The design of voice protective methods cannot be separated from the scenarios to be protected in this paper and the motivation for privacy-preserving, while we also need to characterize theoretically the results that can be achieved by our designed audio protection strategy as well as the problem itself.

\noindent\textbf{Motivation.} Facing malicious speech synthesis and unauthorized data access, effective data defensive methods can better reduce the harm and hazards of the public in the face of attackers. However, on the one hand, with the abuse of technologies such as web crawlers, unauthorized access to large amounts of data is becoming increasingly rampant, which exposes public privacy to a greatly dangerous situation. On the other hand, an effective and generalized voice anti-cloning strategy has not been well proposed. Previous methods have more or less imperfections, such as only deceiving speaker verification models~\cite{wang2023vsmask, huang2021defending} without considering that deepfake speech may not target a specific victim. Therefore, preventing unauthorized abuse of audio and defending against synthetic speech have become the two main motivations of our work.

\noindent\textbf{Unlearnable Audio.}
We assume a clean dataset with $n$ samples to be protected as
$D_c=\left\{(x_i, y_i, z_i)|(x_i, y_i, z_i) \in \mathcal{X} \times \mathcal{Y} \times \mathcal{Z}\right\}_{i=1}^n$, where $x_i$ is the $i$-th speech sample from speaker victim $y_i$ and $z_i$ is the corresponding speech text. Given a TTS model $G$, $G$ can effectively synthesize high-quality and realistic deepfake audio that is similar to the timbre of the specified speaker, after training the speech synthesis model on $D_c$. Users utilize the data protection method we proposed to protect the clean audio dataset $D_c$, adding an imperceptible perturbation $\delta$ to each sample at a protective position $l$ to obtain the protected dataset $D_u$. After training the model $G$ on $D_u$ with the same configuration parameters, the same prompt pair will result in unusable audio filled with background noise at position $l$, thereby achieving the goal of making the protected dataset $D_u$ unlearnable for the TTS model and preserving privacy. Let a speech sample $\mathbf{x}$ to safeguard, the protective method can be specifically described using the following formula:

\begin{equation}
    \begin{aligned}
        \argmin_{\delta}\ & \mathcal{L}(G_l(\mathbf{x}+\delta), \mathbf{x}), \\
        \text{s.t.} \ & H(\mathbf{x}+\delta) \approx H(\mathbf{x}) \quad \text{and} \quad \left\|  \delta \right\|_{p} \le \epsilon,
    \end{aligned}
    \label{eq_definition}
\end{equation}
where $\mathcal{L}(\cdot)$ and $H(\cdot)$ represent the objective and perceptual optimization function reflecting the audio usability respectively. $\left\| \cdot \right\|_p$ measures $\ell_p$ distance of embedded $\delta$ bounded by radius $\epsilon$ for the limitation of human perception. 

Notably, the parameters of $G$ are fixed for perturbation generation, which means using the perturbation to optimize objective function Eq. (\ref{eq_definition}). We employ the gradient-based PGD algorithm~\cite{madry2018towards} to minimize the model's error for better protection.

\subsection{Speaker Selection}\label{section_speaker}
It requires more computing and data resources when training from scratch~~\cite{tajbakhsh2016convolutional}. Based on the pre-training of many samples, it can accelerate and optimize the training process with layers adaptation. For better training, we take into account the degree of voiceprint similarity between the chosen speakers and the pre-trained model of a single speaker during the selection process. Specifically, for the selected speaker denoted as $j$, the similarity between their speaker embeddings can be quantified by cosine similarity, expressed by:
\begin{equation}
d_{j}=D(E_s(x_j), E_s(x_0)),\label{eq_selection}
\end{equation}
where $x_j$ and $x_0$ represent the speech of the $j$-th and targeted speaker, 
and $E_s(\cdot)$ is the speaker encoder that computes the speaker's information features and outputs the embeddings containing personal voiceprint information from the input audio. $D(\cdot)$ computes the cosine similarity of the two vectors.

Meanwhile, we add a limitation that each speaker should own more than 50 samples when selecting speakers to obtain a quantitative balance between different speakers.

\subsection{Unlearnable Audio}\label{section_unlearn_audio}
In the field of computer vision, the imperceptible perturbations generated based on $\ell_p$ norm constraint can make the protected dataset unlearnable for the DNNs~\cite{huang2021unlearnable} in the classification task. The optimization structure is a bi-level loop, with the outer structure for optimizing  model $f_\theta$ and the inner loop for noise generation, which can be written as follows:
\begin{equation}
\operatorname*{argmin}\limits_{\theta} \mathbb{E}_{x,y}[\min_{\delta} \mathcal{L}(f_\theta(x+\delta), y)]\quad
\mbox{s.t.} \quad \left\| \delta \right\|_p \le \epsilon,\label{eq_unlearn}
\end{equation}
where $(x, y)$ denotes the input data and its label.

The bi-level structure will optimize both the model parameters and the embedded perturbations to improve the generalization of the noises across different model parameters. However, this brings a huge time overhead to optimize the model's parameters in large-parameter and large-data-set scenarios such as TTS synthesis. The core of the bi-level error-minimizing is the internal optimization loop, which crafts noise to reduce the error of the model by simulating the training process so that it learns more about the added noise when learning from protected data. Based on the above, we can simplify the bi-level error-minimizing method and use the inner loop as the targeted optimization function which can reduce the cost of training resources. The formulation can be described as
\begin{equation}
\operatorname*{argmin}_\delta  \mathcal{L}(f_\theta(x+\delta), y)\quad
\mbox{s.t.}\quad \left\| \delta \right\| _p\le\epsilon .\label{eq_single}
\end{equation}

Generative TTS models, such as variational autoencoder (VAE), employ an encoder-decoder structure that learns the distribution of input data and then generates a new distribution resembling it. Different from classification problems in the previous researches~\cite{huang2021unlearnable, deng2023v} that the model learns the differences of input data, generative TTS models learn the distribution of input data and are relatively complex about the inputs, such as text, speaker ID, original audio, and spectrogram. We can only apply the perturbation on the clean waveform to ensure that the speech content is not altered. In the process of generating unlearnable examples, we consider the generator $g$ which is the core part of a TTS model for noise generation at position $l$. The simplified objective function, referencing to the Eq. (\ref{eq_single}), can be described as:
\begin{equation}
\operatorname*{argmin}_\delta  \mathcal{L}(g(\mathbf{x}+\delta, \texttt{spec}(\mathbf{x}+\delta), \texttt{text}), \mathbf{x})\quad
\mbox{s.t.}\quad \left\| \delta \right\| _p\le\epsilon , \label{eq_tts}
\end{equation}
where $\texttt{spec}(\cdot)$ computes the linear spectrogram from inputs.

\subsection{Pivotal Objective Perturbation}\label{section_pop}
Due to the diversity of objective functions across various TTS models and our strategy of introducing noise on the waveform to maintain semantic consistency in speech, it is insufficient if we leverage all training objective functions of $g$ as the target for noise optimization. This is because components of the objective function unrelated to input audio would be unable to optimize the noise. Typically, TTS models prioritize the reconstruction function between synthetic and real audio as their primary objective. Moreover, for the generative TTS model, it is conventional to compute the $\ell_1$ (or $\ell_2$) distance between the synthetic and real audio to guide better synthesis which is a learning distribution and outputting it process~\cite{jiang2024megatts, kim2021conditional, kim2020glow, kawamura2023lightweight, shen2018natural, li2019neural}.

For better noise generation, we propose a Pivotal Objective Perturbation approach to synthesize effective protective perturbation. POP selects the \textbf{reconstruction loss} as the objective in the error minimization to generate unlearnable speech samples. We craft SOTA and the backbone TTS model named VITS~\cite{kim2021conditional} as an example to expound the reason and strategy of our method. VITS contains the structure of a decoder and a vocoder, and its generator's objective optimization function can be expressed as follows:
\begin{equation}
\mathcal{L}_{vits}=\mathcal{L}_{recon}+\mathcal{L}_{kl}+\mathcal{L}_{dur}+\mathcal{L}_{adv}(G)+\mathcal{L}_{fm}(G) ,  \label{eq_vits}
\end{equation}
where $\mathcal{L}_{recon}$ denotes the reconstruction loss between real and synthetic speech. $\mathcal{L}_{kl}$ and $\mathcal{L}_{duration}$ represent the KL divergence loss and duration loss. $\mathcal{L}_{adv}(G)$ and $\mathcal{L}_{fm}(G)$ are the adversarial training loss and feature-matching loss of the generator.

In the following, we will give a brief illustration of these loss functions in Eq. (\ref{eq_vits}). KL divergence $\mathcal{L}_{kl}$ is a common loss function of VAE architecture, and it learns the relation between phoneme $c$ and \texttt{text}, \textit{etc.}, which can be expressed as:
\begin{equation}
\begin{aligned}
    \mathcal{L}_{kl}=\log q_\phi\left(z \mid x_{\texttt{lin}}\right)-\log p_\theta\left(z \mid c_{\texttt{text}}, A\right), \\ 
    z \sim q_\phi\left(z \mid x_{\texttt{lin}}\right)=N\left(z ; \mu_\phi\left(x_{\texttt{lin}}\right), \sigma_\phi\left(x_{\texttt{lin}}\right)\right)
\end{aligned}\label{eq_kl}
\end{equation}

The duration loss is a negative variational lower bound, relative to the input text and the length of the text. It can not be  completely affected by speech waveform content, which is formulated as:
\begin{equation}
    \log p_\theta (d|c_{\texttt{text}}) \ge \mathbb{E}_{q_\phi(u,v|d, c_{\texttt{text}})}\left[ \log \frac{p_\theta (d-u, v|c_{\texttt{text}})}{q_\phi(u,v|d, c_{\texttt{text}})} \right].\label{eq_dur}
\end{equation}

$\mathcal{L}_{adv}(G) $ and $\mathcal{L}_{fm}(G) $ are applied for adversarial training. VITS utilizes the discriminator $D$ to distinguish between the output synthetic by the decoder $G$ and the real audio, improving the quality of the synthesized audio. The two functions can be expressed as:
\begin{align}
\mathcal{L}_{adv}(G) & = \mathbb{E}_z\left[(D(G(z))-1)^2\right], \\
\mathcal{L}_{fm}(G)  &= \mathbb{E}_{y, z}\left[\sum_{l=1}^T \frac{1}{N_l}\left\|D^l(y)-D^l(G(z))\right\|_1\right],
\end{align}
where $z$ is the latent variables and $y$ is the ground truth waveform. $T$ denotes the total layers of the discriminator, and $D^{l}$ represents the feature map of layer $l$, having $N_l$ features.

However, the reconstruction loss measures the distance between output and input audio which can be expressed by:
\begin{equation}
    \mathcal{L}_{recon} = ||\texttt{mel}(\mathbf{x}) - \texttt{mel}(\hat{\mathbf{x}})||_1 ,\label{eq_recon}
\end{equation}
where $\mathbf{x}$ and $\hat{\mathbf{x}}$ represent the original and synthesized speech, $\texttt{mel}(\cdot)$ computes the mel-spectrogram from the input and $||\cdot ||_1$ denotes the $\ell_1$ loss function.

For the multi-task learning problem in Eq. (\ref{eq_vits}), we observe that $\mathcal{L}_{dur}$ remains unaffected by perturbations because it is independent of the waveform being altered. Different generative TTS models have distinct objectives. In the VITS model, $\mathcal{L}_{kl}$, $\mathcal{L}_{adv}(G)$, and $\mathcal{L}_{fm}(G)$ can be optimized, but other models, \textit{e.g.}, GlowTTS~\cite{kim2020glow} and Transformer-TTS~\cite{li2019neural}, do not necessarily have them. 
Using them as perturbation optimization functions could result in noise that empirically performs poorly on models that do not include these components. 
Therefore, to ensure high transferability across various models, we can choose a universal pivotal function as the optimization objective that is more effective than optimizing all the losses because it is hard to achieve a complete balance and optimal solution of each hyperparameter by the added perturbation. Moreover, the POP method greatly reduces the computing resources because not all the functions are to be optimized for the real-world scenario's application and generative TTS models also output the waveform to compute Eq. (\ref{eq_recon}). We will explain more reasons for selecting $\mathcal{L}_{recon}$ for perturbation optimization in Section \ref{section_exp_ablation} in the aspect of protective effect utilizing each loss function. 

High-quality speech synthesis requires large-scale datasets for training, so some advanced TTS models are trained using an approach named windowed generator training (WGT)~\cite{kim2021conditional, kawamura2023lightweight}, where small portions of a speech are utilized for training at each iteration, speeding up the training process and improving the model's generalization. if we craft perturbations for the whole audio segment, it will bring more computational overhead and improve the perceptibility. Therefore, for the model employing WGT (\textit{e.g.}, VITS~\cite{kim2021conditional} and MB-iSTFT-VITS~\cite{kawamura2023lightweight}), we consider a position-based strategy in POP, referring to PosCUDA~\cite{gokul2024poscuda}, which generates noise by an audio patch at position $l$, (\textit{e.g.}, $\mathbf{0}$), so that only a certain patch is perturbed, which is less time-consuming and more imperceptible. If the model is lightweight and uses the whole audio for training instead of WGT (\textit{e.g.}, GlowTTS~\cite{kim2020glow}), POP protects against the entire audio, which achieves better protection. We will discuss in detail the effect of protecting against a position-based patch and entire audio under different training methods in Section \ref{section_exp_robust}.

\section{Experiments and Analyses}
In this section, we outline the datasets and experimental details. Furthermore, we present and discuss the anti-cloning effectiveness, transferability, and robustness across different SOTA models.

\subsection{Experimental Settings}\label{section_exp_set}
\noindent\textbf{Datasets.} We utilize two famous and clean speech datasets, LibriTTS~\cite{zen2019libritts} and CMU ARCTIC~\cite{kominek2003cmu}. For the LibriTTS dataset, we select the top 50 speakers that are the most similar to the pre-trained speaker from the train-clean-100 subset derived from the LibriSpeech~\cite{panayotov2015librispeech} corpus and use ECAPA-TDNN~\cite{desplanques20_interspeech} as speaker encoder. The CMU ARCTIC dataset comprises 18 speakers, from each of whom we select 300 samples for fine-tuning. During fine-tuning, 80\% of the samples are used for training randomly, while 20\% are reserved for evaluation.

\noindent\textbf{Models.} The experimental models selected show SOTA performance in the TTS domain. These models include:
\begin{itemize}[leftmargin=1em, topsep=0pt]
    \item GlowTTS~\cite{kim2020glow}: GlowTTS is a two-stage speech synthesis model that learns to synthesize mel-spectrograms as a synthesizer;
    \item VITS~\cite{kim2021conditional}: VITS is a backbone TTS model based on the VAE structure in the speech system domain. It operates as an end-to-end synthesis system, learning directly from the input waveform to produce similar waveform distributions;
    \item MB-iSTFT-VITS~\cite{kawamura2023lightweight}: MB-iSTFT-VITS is an enhanced VITS model utilizing multi-band inverse Short-Time Fourier Transform which makes the model efficient.
\end{itemize}
For GlowTTS, we choose WaveGlow~\cite{prenger2019waveglow} and HiFiGAN~\cite{kong2020hifi}, the SOTA processor conversing mel-spectrogram to waveform, respectively as a vocoder. Three models are trained on the LJSpeech~\cite{ljspeech17} dataset containing samples from a single speaker. 

\noindent\textbf{Experimental Details.} 
We keep the conventional settings as in previous papers~\cite{kim2020glow,kim2021conditional,kawamura2023lightweight} and perform position-fixed cropping during noise generation and training procedure with a batch size of 15. To ensure the effectiveness of fine-tuning, we set the number of training iterations to 200, with 200 iterations for noise generation. We set the noise boundary $\epsilon$ as 8/255 to strike a balance between enhancing the anti-cloning protective effect and maintaining human perceptibility about the embedded noise. The experiments are conducted on one NVIDIA A800 GPU with 80GB memory.

\begin{table*}[t]
  \centering
  \caption{The protective effectiveness comparison across different TTS models trained on clean, random noise added, error-minimizing (EM) noise and pivotal objective perturbation (POP) protected dataset objectively with MCD and WER metrics on the LibriTTS dataset $D_1$ and the CMU ARCTIC dataset $D_2$. }
    \begin{tabular}{ccccc cccccccccc}
    \toprule
    \multirow{2}[4]{*}{Dataset} & \multirow{2}[4]{*}{Method}  
    & \multicolumn{2}{c}{MB-iSTFT-VITS} & \multicolumn{2}{c}{VITS}& \multicolumn{2}{c}{GlowTTS$^{\mathrm{a}}$} & \multicolumn{2}{c}{GlowTTS$^{\mathrm{b}}$}  \\
    \cmidrule(r){3-4}\cmidrule(lr){5-6}\cmidrule(lr){7-8}\cmidrule(l){9-10}
    & & MCD($\uparrow$)   & WER($\uparrow$)
    & MCD($\uparrow$)   & WER($\uparrow$)
    & MCD($\uparrow$)   & WER($\uparrow$)
    & MCD($\uparrow$)   & WER($\uparrow$)\\
    \midrule
    \multirow{5}{*}{$D_1$}
        & ground truth    & - & 13.954 & - & 13.954 & -& 13.954 & - & 13.954 \\
        & clean & 5.830  & 21.939  & 5.791 & 27.033  
            & 7.597 & 31.672  & 7.852 & 30.998       \\
        & random noise & 6.019 & 32.618  & 6.299  & 41.828 
            & 8.437 &39.895   &9.591 &36.629      \\
        & EM~\cite{huang2021unlearnable}
            & 11.463 & 99.681 & 10.173 & \textbf{116.091}
            & 15.206 & 102.572 & 17.529 & 98.687\\
        \cmidrule(lr){2-10}
        & \textbf{POP (ours)} &\textbf{13.646} &\textbf{127.310} & \textbf{13.440} & 105.596
        &\textbf{15.237} &\textbf{109.766}    & \textbf{17.563} &\textbf{100.713} \\
    \midrule
    \multirow{5}{*}{$D_2$}
        & ground truth & -  & 12.474   &- &12.474  &- & 12.474   & - & 12.474 \\
        & clean &6.125 &18.976  & 6.239  & 28.381  
            & 7.443  &34.794    & 7.624 &33.610   \\
        & random noise &6.820  &29.131 & 6.617  & 29.879 
            &8.173 &41.882  &9.490 &38.973         \\
        & EM~\cite{huang2021unlearnable}
            & 10.557 & 59.301 & 13.432 & 97.085 
            & 12.265 & 91.389 & 14.036 & \textbf{88.046} \\
        \cmidrule(lr){2-10}
        & \textbf{POP (ours)} &\textbf{13.178} &\textbf{97.685}  
            &\textbf{14.086}  &\textbf{100.756} 
            &\textbf{12.288}  &\textbf{92.273}  
            &\textbf{14.149}  & 84.528      \\
    \bottomrule
    \multicolumn{4}{l}{$^{\mathrm{a}}$ Mel+WaveGlow, $^{\mathrm{b}}$ Mel+HiFiGAN.}
    \end{tabular}
  \label{tab:1}
\end{table*}

\noindent\textbf{Metrics.} To comprehensively evaluate the effectiveness of training on both clean and protected datasets, we consider objective metrics such as mel-cepstral distortion (MCD) with dynamic time warping and word error rate (WER) \text{(\%)}, along with subjective evaluation using mean opinion score (MOS) with 95\% confidence intervals. For objective assessment, WER reflects the recognizability of pronunciation in a given speech, utilizing the pre-trained medium size of Whisper~\cite{radford2023robust}, an open-source speech recognition model from OpenAI. MCD can reflect the differences in terms of timbre, speech content, and duration time between the generated and synthetic speech. For the subjective metric, MOS denotes the most intuitive reflection of the usability of these deepfake speeches because malicious speech synthesis aims to spoof human perception by marking a 0 to 5 score for each audio. The higher value of WER represents more unclarity of a speech, and the higher value of MCD reflects less-than-ideal training results. Moreover, for speech imperceptibility, we consider the signal-to-noise ratio (SNR)(dB) and perceptual evaluation of speech quality (PESQ) which measure the perturbation and original audio. Higher SNR represents lower noise disruption with better speech clarity the same as PESQ.

\noindent\textbf{Baselines.} Previous voice protection methods~\cite{yu2023antifake, wang2023vsmask} based on adversarial examples usually aim at spoofing speaker verification systems with high-quality and clear speech, which does not fully align with our method. Therefore, for stronger adaptive protection at train time, we select a great effective approach in perturbative availability poison (PAP) attacks~\cite{liu2023image}, error-minimizing (EM)~\cite{huang2021unlearnable}, to generate specific perturbation. Moreover, to evaluate the effect of the non-specific algorithm for perturbation generation, we consider adding random Gaussian noise with the same constraint radius as another baseline.

\subsection{Effectiveness of POP Method}\label{section_exp_effect}
In this section, to verify the protective effectiveness of our method, we conduct experiments on two large-scale multi-speaker datasets with each of the three TTS models as we mentioned previously. The evaluation involves comparing ground truth with generated synthesized audio by different methods using the same sentence spoken by the same speaker.

We process these training samples in different ways. First, we use a clean dataset without perturbation to verify the model's outstanding speech synthesis performance. In addition, we utilize baseline methods and our proposed data protection method POP to generate specific perturbations for each TTS model and evaluate the results when training on the protected dataset.

Table \ref{tab:1} presents our experimental results. It shows that models perform well with relatively moderate fine-tuning on the selected datasets. For instance, the WER of samples generated by the MB-iSTFT-VITS model differs from ground truth by only 6.5\%, which is also reflected in subjective evaluations in Section \ref{section_user}, indicating the effectiveness of our fine-tuning approach. When random noises are added, the synthesized audio by the model exhibits a small number of noises resembling background sounds, yet the articulation and clarity of the speaker's pronunciation remain comparable to clean data. For example, the synthesized speech quality difference is negligible for VITS trained on the clean and randomly noisy CMU ARCTIC dataset, suggesting that random noise does not render the data samples unlearnable. The random noise-added dataset shows that undesigned perturbations cannot interfere with the model to achieve data protection.

As shown in Table \ref{tab:1}, the audio data protected by our proposed POP data protection method can be employed on different TTS models and achieve a remarkably effective anti-cloning effect. On the LibriTTS dataset, the WER and MCD of the MB-iSTFT-VITS model increase significantly from 5.830\% and 21.939\%  when trained on clean samples to 13.646\% and 127.310\% trained on POP-protected dataset, indicating that the model merely learns any relevant voice timbre and text alignment information from the protected waveform. The synthesized audio is almost entirely noise without relevant audio information from an auditory perspective. At the same time, similar effects are observed on other models and datasets. The experiments in this section demonstrate that the POP method has significant effectiveness in preventing audio data from high-quality malicious voice cloning.

\begin{table*}[t]
  \centering
  \caption{The results of MOS subjective evaluation about synthetic audio across different datasets and methods.}
  \resizebox{0.95\linewidth}{!}{
  \begin{tabular}{ccccc ccccc}
    \toprule
    \multirow{2}[1]{*}{Method}
    & \multicolumn{4}{c}{LibriTTS} & \multicolumn{4}{c}{CMU ARCTIC} \\
    \cmidrule(r){2-5}\cmidrule(lr){6-9}
    & MB-iSTFT-VITS &VITS & GlowTTS$^{\mathrm{a}}$ & GlowTTS$^{\mathrm{b}}$
    & MB-iSTFT-VITS &VITS & GlowTTS$^{\mathrm{a}}$ & GlowTTS$^{\mathrm{b}}$
    \\
    \midrule
    ground truth & 4.71$\pm$0.21 & 4.71$\pm$0.21 & 4.71$\pm$0.21 & 4.71$\pm$0.21
                 & 4.66$\pm$0.15 & 4.66$\pm$0.15 & 4.66$\pm$0.15 & 4.66$\pm$0.15\\
    clean & 4.50$\pm$0.25 & 3.68$\pm$0.45 & 4.74$\pm$0.17 & 4.69$\pm$0.19 
          & 4.30$\pm$0.35 & 3.39$\pm$0.39 & 3.33$\pm$0.39 & 3.97$\pm$0.32\\
    random noise & 3.44$\pm$0.31 & 2.79$\pm$0.37 & 3.08$\pm$0.46 & 3.71$\pm$0.35
                 & 2.97$\pm$0.48 & 3.21$\pm$0.39 & 2.66$\pm$0.48 & 1.58$\pm$0.43\\
    EM~\cite{huang2021unlearnable} 
       & 0.45$\pm$0.14 & 0.18$\pm$0.14 & 0.85$\pm$0.24 & 1.45$\pm$0.31
       & 1.58$\pm$0.42 & 1.33$\pm$0.34 & \textbf{1.43$\pm$0.39} & 1.45$\pm$0.48\\
    \midrule
    \textbf{POP (ours)} & \textbf{0.18$\pm$0.13} & \textbf{0.13$\pm$0.09} 
                        & \textbf{0.78$\pm$0.21} & \textbf{0.92$\pm$0.27}
                        & \textbf{0.65$\pm$0.31} & \textbf{0.54$\pm$0.24} 
                        & 1.80$\pm$0.50 & \textbf{1.45$\pm$0.38}\\
    \bottomrule
  \end{tabular}
  }
  \label{tab_mos}
\end{table*}

\subsection{Transferability Analyses}\label{section_exp_transfer}
When generating noise using different models, we employ the same model in fine-tuning and noise generation. However, there are various advanced TTS models. Therefore, the question arises: 

\begin{tcolorbox}
  \begin{center}
    \textit{Can the unlearnable samples produced by one surrogate model be effective on other models as well?}
  \end{center}
\end{tcolorbox}

In this section, we utilize MB-iSTFT-VITS as the surrogate model to protect datasets. Then we train on two other models using surrogate model-protected audio to verify the transferability and generalization of the specific perturbation across even unseen models.

\begin{table}[t]
  \centering
  \caption{Transferability of the noises from MB-iSTFT-VITS.}
  \begin{tabular}{ccccccc}
    \toprule
    \multirow{2}[2]{*}{Model} & \multirow{2}[2]{*}{Model} 
    & \multicolumn{2}{c}{LibriTTS} & \multicolumn{2}{c}{CMU ARCTIC}\\
    \cmidrule(r){3-4}\cmidrule(lr){5-6}
    & & MCD($\uparrow$)   & WER($\uparrow$)
    & MCD($\uparrow$)   & WER($\uparrow$) \\
    \midrule
    \multirow{3}[2]{*}{EM}
        & VITS &12.935 &106.829 &13.082 &105.611  \\
        & GlowTTS$^{\mathrm{a}}$ &11.909 &74.492 &10.429 &74.137  \\
        & GlowTTS$^{\mathrm{b}}$ &10.673 &64.613 &11.886 &72.816 \\
    \midrule
    \multirow{2}[2]{*}{\textbf{POP}}
        & VITS &13.218 &124.013 &13.899 &101.590  \\
        & GlowTTS$^{\mathrm{a}}$ &13.829 &81.110 &9.342 &68.277  \\
        & GlowTTS$^{\mathrm{b}}$ &16.258 &75.862 &10.675 &66.521 \\
    \bottomrule
  \end{tabular}
  \label{tab_transfer}
\end{table}

Table \ref{tab_transfer} presents the protective performance of the noise generated by the surrogate model and transferred to other models. Surprisingly, although the procedure of perturbation generation does not rely on the corresponding training model and the structures of TTS models are various, our method still achieves great protection effects. It is noteworthy that the audio protection of the surrogate model in the VITS model achieves better results than self-protection. On the LibriTTS dataset, the WER value of utilizing the POP method is 124.013\%, which is higher than the self-protection value of 105.596\%. This reflects the high transferability of our method. Moreover, GlowTTS and MB-iSTFT-VITS have significant differences in terms of model structure and optimization objectives but can achieve good transferability. The WER on the LibriTTS dataset is 81.110\%, and most of the synthesized audio is inaudible, greatly preserving personal privacy information from being synthesized.

These findings fully illustrate the transferability of our proposed data protection method when employing a surrogate model to generate perturbation. In the real world, we cannot predict the model the adversary used when training. The highly migratory data protection method can be a much more effective approach to solve the potential malicious speech synthesis attack caused by the lack of \textit{a prior} knowledge about the model trainer.

\subsection{User Study}\label{section_user}
In Sections \ref{section_exp_effect} and \ref{section_exp_transfer}, we performed a comprehensive objective verification study of our proposed method using three distinct TTS models. A crucial aspect of the method evaluation was examining the human perception of deepfake speeches generated by TTS models when training on different perturbation-embedded datasets. For each protection method, we randomly selected two audio samples and created a questionnaire via the Credamo platform. We invited 73 participants to complete the questionnaire, ultimately obtaining 61 valid responses. To best reduce subjective biases, we implemented three key measures: (1) anonymizing the purpose of the questionnaire and the names of the audio samples, (2) including two arithmetic questions to ensure participants' attentiveness and (3) randomizing the order of the samples to prevent the order of the samples from having an impact on participant ratings. Finally, we calculated the subjective metric MOS with 95\% confidence intervals, following the principles outlined in \cite{kumar2019melgan}.

Table \ref{tab_mos} shows the subjective experimental results across three TTS models and various methods. From the results in Table \ref{tab_mos}, it can be found that unprotected clean data can easily be exploited to synthesize high-quality realistic audio. On the LibriTTS dataset, the subjective evaluation result of MB-iSTFT-VITS is 4.50$\pm$0.25, which is almost comparable to the 4.71$\pm$0.21 of real audio. Unprotected data brings the owner of the audio into a potentially dangerous area, causing leakage of sensitive data such as voiceprint. On the contrary, employing POP to protect the dataset cannot synthesize audio that deceives the human ear perception, greatly reducing the risk of data being in a dangerous area. On the VITS model, its MOS value is only 0.13$\pm$0.21, meaning that the synthesized audio is just the noisy audio content to the human ear and they will not be deceived by these ``bad'' audio. This demonstrates that our approach is delivering significant results in securing data and effectively preventing the leakage of sensitive information.

\subsection{Ablation Study}\label{section_exp_ablation}
In this section, we conduct the ablation study on the selection of different objective functions and random segment training with a supplemental defensive strategy to improve the protection effect in this scenario. Moreover, we use Eq. (\ref{eq_selection}) to select and conduct this experiment on the single speaker most similar to the pre-trained speaker, which will be detailed in  Section \ref{section_exp_robust}.

\noindent\textbf{Objective Analyses.} In Section \ref{section_pop}, we have introduced the challenges when using the perturbative protection strategy, and regard a backbone model VITS as an example to illustrate the reasons for pivotal function selection and the detailed information about POP strategy. Our method POP selects $\mathcal{L}_{recon}$ in Eq. (\ref{eq_vits}) as the optimization objective, because choosing this function can bring about the best protection effect, and generally speaking, generative TTS models usually regard the reconstruction loss as one of the objectives in multi-objective optimization. In this experiment, we will explore the protective effects brought by selecting other functions in Eq. (\ref{eq_vits}) as the target objective, specifically: \circled{a} $\mathcal{L}_{recon}$, \circled{b} $\mathcal{L}_{kl}$, \circled{c} $\mathcal{L}_{dur}$, \circled{d} $\mathcal{L}_{adv}(G)$, \circled{e} $\mathcal{L}_{fm}(G)$.

\begin{figure}[t]
\centerline{\includegraphics[width=0.30\textwidth]{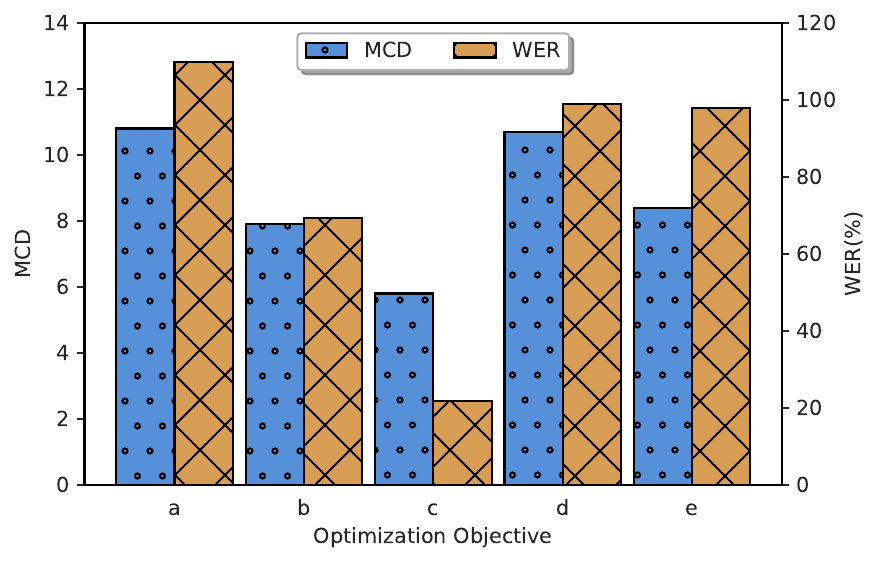}}
\caption{Ablation study on the protection effect with different optimization objectives.}
\label{fig_ablation}
\end{figure}

Figure \ref{fig_ablation} shows the protection effect of the generated perturbation when utilizing different objective functions. Since function \circled{c} cannot be affected by perturbation, we consider that the generated noises are zero vectors via \circled{c}. From the figure, we can observe that employing function \circled{d} or \circled{e} can achieve great protection results, while other generative TTS models may not have this loss function, which cannot guarantee high transferability across unknown models, such as GlowTTS lacking these two functions. However, on the one hand, the function \circled{a} realizes the highest protection effect among these five objects. On the other hand, the perturbation comes from the measure between the generated speech and the original one empirically aligns with the task of generative TTS, ensuring that the perturbation remains effective across different TTS models. Therefore, we select function \circled{a} as the optimization objective of the POP method.

\noindent\textbf{Random Segment Training.} 
In Eq. (\ref{eq_definition}), for the training strategy of random WGT, we consider protecting speeches at the position-fixed patch, which not only reduces the computing resources and time cost but also enhances the perturbation imperceptibility. In the previous experiments, our training strategy focuses on an assumption that the adversary trains the protected samples at position $l$. However, a more realistic scenario is that the adversary can conduct an attack on a random segment (RS), which necessitates a stronger defense that the perturbation can cover the entire audio against an RS attack scenario. Therefore, we consider two supplemental perturbative protection methods based on POP: random segment perturbation (RSP) and entire segment perturbation (ESP). The coming out of RSP is inspired by \textit{expectation over transformation}~\cite{athalye2018synthesizing}, a robust adversarial examples generation mechanism, to synthesize noise at random position for each optimization iteration instead of a fixed position. Based on this, we can generate a better effective perturbation against RS, while the variance of each optimization is large. ESP is directly optimizing the entire segment of audio.

\begin{table}[t]
  \centering
  \caption{The protective performance when training the VITS model utilizing RS (or not) across different methods. ``time'' represents the average time cost for protecting a sample speech with 200 iterations on VITS.}
  \resizebox{0.95\linewidth}{!}{
  \begin{tabular}{ccccc ccccc}
    \toprule
    \multirow{2}[1]{*}{Method} & \multirow{2}[1]{*}{time(s)($\downarrow$)}
    & \multicolumn{2}{c}{w/ o RS} & \multicolumn{2}{c}{RS} \\
    \cmidrule(r){3-4}\cmidrule(lr){5-6}
    & & MCD($\uparrow$)   & WER($\uparrow$)
    & MCD($\uparrow$)   & WER($\uparrow$) \\
    \midrule
    clean & - & 5.526 & 25.703 & 5.279 & 26.555 \\
    EM & 2.154 & 10.653 & 96.988 & 7.696 & 57.004 \\
    \midrule
    POP             & \textbf{1.855} & 10.838 
                    & \textbf{109.571} & 7.659 & 63.021 \\
    POP+\textbf{RSP}    & 1.880 & 10.259 
                        & 74.520 & 10.101 & 61.400 \\
    POP+\textbf{ESP}    & 7.128 & \textbf{11.144} & 84.683 
                        & \textbf{10.294} & \textbf{89.930}\\
    \bottomrule
  \end{tabular}
  }
  \label{tab_rs}
\end{table}

Table \ref{tab_rs} shows the superiority of the POP method facing training without RS, with great protection effect in maintaining low time overhead conditions. When the more realistic RS method is employed, the POP-protected dataset is still somewhat protective, with the synthesized speech having a higher speech unclarity score (WER)  than that of ``POP+\textbf{RSP}''.

\begin{table*}[t]
  \centering
  \caption{The robustness quantization via various speech augmentation and defensive methods from 1 to 13  on our proposed method POP evaluated by MCD and WER metrics. The \underline{underline} values indicate the most significant decreases in protection compared to training without speech augmentation (``w/ o'' in the Table) .}
  \resizebox{0.98\linewidth}{!}{
  \begin{tabular}{ccccc ccccc ccccc ccccc}
    \toprule
    \multirow{2}[2]{*}{Method} & \multirow{2}[2]{*}{Metric}
    & \multirow{2}[2]{*}{w/ o} & \multicolumn{3}{c}{Adversarial Defender}
    & \multicolumn{7}{c}{Audio Processor} & \multicolumn{3}{c}{Filters}\\
    \cmidrule(l){4-6}\cmidrule(lr){7-13}\cmidrule(r){14-16}
    & & & \circled{1} & \circled{2} &\circled{3} 
    & \circled{4} & \circled{5} &\circled{6}
    & \circled{7} & \circled{8} &\circled{9}
    & \circled{10} & \circled{11} &\circled{12} &\circled{13}\\
    \midrule
    \multirow{2}{*}{\textbf{POP}}
        & MCD($\uparrow$)   & 10.838 & 8.232 & 11.097 & 10.549 & 11.056 
                            & 11.275 & 7.989 & 10.550 & 9.941 & 8.013
                            & 7.392 & 11.255 & 11.163 & \underline{7.284}\\ 
        & WER($\uparrow$)   & 109.571 & 108.390 & 102.861 & 103.570 & 109.851 
                            & 98.582 & 101.079 & 100.876 & 106.710 & \underline{88.009}
                            & 98.342 & 98.060 & 97.923 & 90.317 \\
    \bottomrule
  \end{tabular}
  }
  \label{tab_augmentation}
\end{table*}

\subsection{Robustness against Adaptive Attackers}\label{section_exp_robust}
In the real world, we cannot predict the training methods that various adaptive attackers will adopt. A stronger adversary can keenly perceive the difference between audio protected by the POP method and unprotected clean audio. Therefore, they can use various attack methods targeting perturbations, with perturbation removal technology and data augmentation technology being the most commonly used. In the previous experiments in Sections \ref{section_exp_effect}, \ref{section_exp_transfer}, and \ref{section_user}, we utilize large-scale multi-speaker datasets for evaluation, typically with thousands of training samples. There is a more realistic scenario that attackers cannot access such a large number of samples. Therefore, we select a single speaker with the closest voiceprint similarity to the pre-trained speaker, containing 117 samples that only last 16 minutes, as used in Section \ref{section_exp_ablation} to simulate the most challenging protection scenarios. In this section, we utilize the VITS model to conduct noise reduction and speech augmentation techniques.

\noindent{\textbf{Noise Reduction.}} From the experiment in Section \ref{section_exp_effect}, it can be found that the addition of perturbations to clean samples, whether generated specifically or randomly, will have a certain weakening effect on the TTS models. When an attacker obtains audio protected by POP, the experienced attackers will detect the embedded abnormal perturbations and become suspicious of the obtained training samples. In the audio field, their most effective way is to directly use perturbation removal methods on the added perturbations to eliminate the protection effect caused by noise. However, due to the lack of clean samples from the adversary, they cannot obtain unprotected samples based on our public perturbation generation methods. Usually, they employ perturbation removal techniques in the audio field. In this experiment, we simulate the adversary using the specific noise reduction technique~\cite{tim_sainburg_2019_3243139} using spectral gating which computes the spectrogram of a speech and estimates a perturbation threshold for each frequency band to denoise. 

After applying the noise reduction method to denoise the samples protected by POP, the resulting samples do not contain a noisy background, and the effect of the perturbations we embedded has been highly destroyed. When training the VITS model using the audio after denoising with noise reduction, the values of MCD and WER are 8.699 and 83.419\% respectively compared to 10.838 and 109.571\% without the noise reduction. This indicates that after denoising the embedded perturbation, the POP method still maintains a high level of protection. This, in turn, validates that after publishing audio protected by POP, adversaries cannot easily restore the original audio, thereby enhancing the preservation of our privacy.

\noindent{\textbf{Speech Augmentation.}} 
In the real world, attackers can also employ a series of data augmentation methods when training TTS models. Previous unlearnable examples are vulnerable to data augmentation~\cite{huang2021unlearnable, fu2022robust} and cannot resist the structural damage caused by various specific data transformations. We cannot entirely predict how the audio after uploading by the users will transform when transmission, upload, and download, \textit{etc.}, as well as to the speech augmentation employed by adaptive attackers. Therefore, our proposed data protection method should be robust against various speech augmentation techniques. 

In this experiment, we consider three different series of speech augmentation techniques: Adversarial Defender, Audio Processor, and Filters. In Adversarial Defender, we utilize an audio-defensive transformation against adversarial examples proposed by~\cite{huang2021defending} to test the performance of the POP method in the face of specifically designed techniques. Specifically, we employ \circled{1} Down Sampling and Up Sampling, \circled{2} Mel-spectrogram Extraction and Inversion, \circled{3} Quantization and Dequantization techniques. In Audio Processor, we simulate audio transformations and compression techniques that may be used in daily life including \circled{4} Speed Adjustment, \circled{5} Adding Gaussian Noise, \circled{6} Time Mask, \circled{7} Pitch Shift, \circled{8} MP3 Compression, \circled{9} Hybrid Transformation (combining five techniques \circled{4} $\to$ \circled{8} for a stronger structural destruction), and we also consider an advanced spectrogram masking technique that is greatly used and effective in speech augmentation \circled{10} Spec Frequency Mask~\cite{park19e_interspeech}. Moreover, in Filters, we consider three filtering techniques that are more effective against perturbations \circled{11} Band-Pass Filter, \circled{12} High-Pass Filter, \circled{13} Low-Pass Filter.~\footnote{We employ \url{https://github.com/iver56/audiomentations} to conduct Audio Processor and Filters. More detailed speech transformation is shown in the Appendix~\ref{section_appendix}.}

The experimental results are shown in Table \ref{tab_augmentation}. From this table, we can find that for audio protected by POP, the above speech transformation techniques, when compared to not using them, exhibit a certain degree of weakening effect on the protection, while the effect is relatively limited. After utilizing Low-Pass Filter on the protected audio, the values of MCD and WER decrease to 7.284 and 90.317\% respectively, which can reduce the interference caused by embedded perturbation on the model to a certain extent. In addition, the most efficient data transformation can reduce MCD and WER to 7.284 and 88.009\%, respectively. However, compared to training on the original samples, the exploitation of data augmentation techniques still cannot completely disrupt the structure of the perturbations. In other words, the POP method remains highly robust despite the advanced and most effective data augmentation techniques in the audio field.

\section{Discussions and Limitations}\label{section_exp_discussion}
In this section, we outline some discussion points about the detailed information and limitations of our proposed method.


\begin{figure}[t]
       \centering
       \subcaptionbox{SNR values.}{\includegraphics[width = 0.22\textwidth]{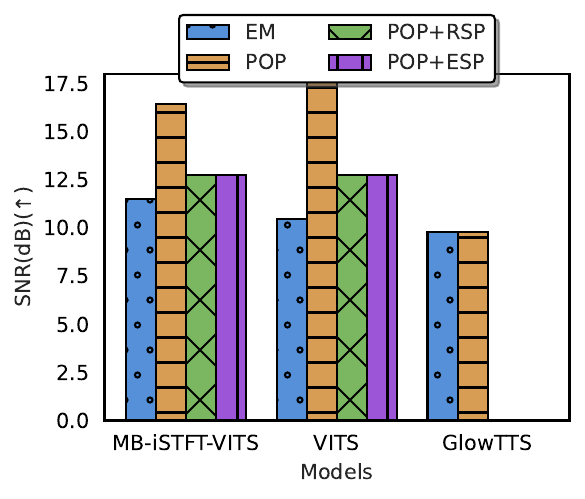}\label{fig_perception_a}}
       \subcaptionbox{PESQ values.}{\includegraphics[width = 0.22\textwidth]{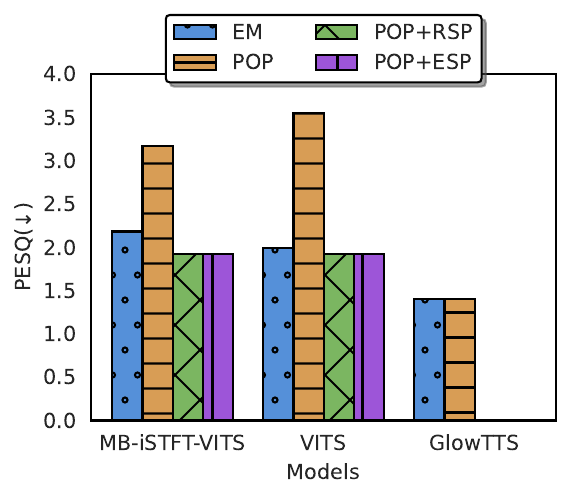}\label{fig_perception_b}}
       \subcaptionbox{Mel-spectrogram comparison.}{\includegraphics[width = 0.44\textwidth]{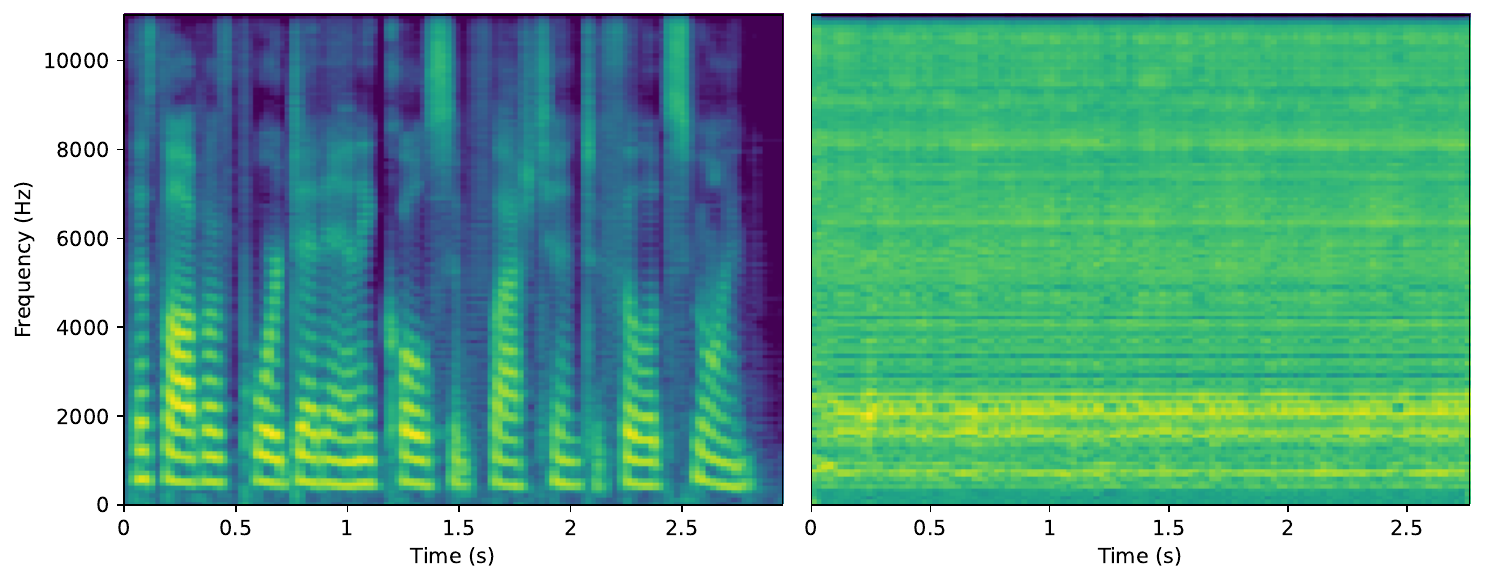}\label{fig_perception_c}}
       \caption{Audibility comparison on the protected dataset.}
   \end{figure}

\noindent\textbf{Audibility Analyses.} The perturbation we embedded on the original audio should satisfy a better imperceptibility to guarantee audio usability in realistic scenarios and reduce the adversary's alertness to detect anomalously embedded noise. Figure \ref{fig_perception_a} and \ref{fig_perception_b} shows the perceptibility of the perturbation for the three models, where the POP method provides the best protective effect, with the SNR and PESQ of 17.894 and 3.551, respectively, in the VITS model, which represents high audio usability. The POP approach causes the least change to the dataset and has the highest imperceptibility compared to the EM approach and whole-segment audio protection.

In addition, the added noises are constrained by the $\ell_p$ norm strictly so that the maximum value of the perturbation does not exceed the radius $\epsilon$, limiting the perturbation interference effect with the audio. We calculate the mean square error (MSE) between the original and protected audio to be 99\%, which means that the two audio waveforms are greatly similar and not significantly different. Figure \ref{fig_perception_c} shows the mel-spectrogram comparison of the synthesized speech when training on clean samples and the POP-protected dataset. It can be noticed that the left one contains a large amount of audio information while the right one has almost no content embodied in it, reflecting that the synthesized audio does not reveal private and sensitive information.

\noindent\textbf{Subjective Biases.~}In the real world, synthetic speech often requires deception to human auditory perception. Therefore, in this paper, an important aspect of achieving audio data protection is to consider whether the human ear can be deceived by the synthesized audio. Non-objective evaluations conducted on humans are often accompanied by subjective biases of the participants themselves. For example, if a participant knows the content and intention of the project in advance, he may be more inclined to participate in the direction he prefers, and it may also be related to the participation time, mood, \textit{etc.}, thus reducing the credibility of the obtained results. Therefore, to minimize potential biases, we completely anonymize the title of our questionnaire and the audio file name, while also adding test sample questions to detect whether the participant is paying attention or not. To improve the credibility of subjective results, we refer to the method~\cite{kumar2019melgan} and calculate a 95\% confidence interval to fully reflect our test results.

\noindent\textbf{Robustness Test.~}In previous experiments, our method has demonstrated remarkable effectiveness and transferability across different TTS models. In real-world protection, the embedded perturbations need to withstand various potential transformation methods, so the robustness guarantee of the generated noise is also a crucial aspect. The data privacy preservation method used in this paper is a special type of clean-label and triggerless data poisoning attack. Currently, there are some defense methods against data poisoning, but they are specially designed for classification tasks, measuring the most effective defense methods in the categorical space. However, this paper focuses on the generative task, where the key lies in the generation of realistic deepfake samples, not the label domain. Therefore, many previous defense methods~\cite{yang2022not, borgnia2021dp} cannot be directly applied. Currently, there is a lack of effective defense methods against attacks in the audio field, which is almost stagnant. Based on this,  we verify the effectiveness and transferability of the method and can resist simple data augmentation transformations.

\noindent\textbf{Ethical Considerations.} The subjective evaluation has obtained the consent of the participants. All participants we have invited are over 18 years old without collecting additional private information.

\section{Conclusion}
In this paper, we focus on the defense mechanism against unauthorized exploitation of audio samples. It may bring huge threats in the real world when uploading our sensitive information on social platforms without protection. To cope with this and achieve privacy-preserving, we devise an effective and transferable perturbative data protection method named Pivotal Objective Perturbation. It aims to make training samples unlearnable by applying specific imperceptible error-minimizing perturbation. Compared to the EM method in PAP, our proposed POP can not only reduce the computing resources but also achieve a better data protection effect because not all the functions can be optimized by the perturbation and the pivotal optimization can reduce the model error better. We validate our method's effectiveness on common datasets with advanced TTS models. Subjective and objective evaluations have shown their outstanding validity in protecting audio data. The generated noise transferred well across models, showing our method's improvement. Moreover, our method remains highly robust against adaptive adversaries. Compared to existing works, our approach offers a convenient, novel, and transferable audio protection mechanism in the TTS domain by rendering samples unlearnable. This represents a step toward the security and privacy of audio data, mitigating the potential risks associated with voice cloning.

\section{Acknowledgement}
We thank the reviewers for their insightful and helpful feedback on our work. This work was supported in part by the National Key Research and Development Program of China under Grant No. 2020YFB1805400 and the Fundamental Research Funds for the Central Universities under Grant No. 2024ZCJH05.

\bibliographystyle{ACM-Reference-Format}
\balance
\bibliography{egbib}

\appendix
\section{Detailed Information}\label{section_appendix}
In this section, we illustrate the detailed descriptions of speech transformation techniques conducted in Section \ref{section_exp_robust} by Table \ref{tab_trans_describe}.

\begin{table*}[b]
  \centering
  \caption{The descriptions of each speech transformation technique.}
  \renewcommand{\arraystretch}{0.3}
  \resizebox{0.9\linewidth}{!}{
  \begin{tabular}{ccc}
    
    \toprule
    Number & Speech Transformation & Description and Parameter Settings\\
    \midrule
    \circled{1} & Resample & \makecell[l]{ 
        We first downsample the input waveform to a random one among 8k, 10k, or 12k, and\\
        subsequently, we upsample it back to its original sampling rate. } \\ \\
    \circled{2} & Mel-spectrogram Inversion & \makecell[l]{
        We extract the mel-spectrogram from the input original audio and convert it back to \\
        the original audio, albeit with some loss of the original audio signals.} \\ \\
    \circled{3} & Quantization & \makecell[l]{
        As an effective defense method in audio transformation, we first quantize the audio to\\
        8 bits and then painstakingly reconstruct it to its original form.} 
        \\ \\
    \circled{4} & Speed Adjustment & \makecell[l]{
        For any given audio input, we randomly select a speed coefficient from 0.8, 0.9, 1.0, 1.1, \\
        or 1.2, apply it and then restore the audio to its original speed.} \\ \\
    \circled{5} & Gaussian Noise & \makecell[l]{
        We add some Gaussian noise to a speech recording within the radius.} \\ \\
    \circled{6} & Time Mask & \makecell[l]{
        Some portions (10\% to 15\%) of the audio clips are randomly muted.} \\ \\
    \circled{7} & Pitch Shift & \makecell[l]{
        This transformation randomly adjusts the pitch of the audio within a range of -4  to +4\\ 
        semitones, with a 50\% chance of actually being applied.} \\ \\
    \circled{8} & MP3 Compression & \makecell[l]{
        An MP3 encoder is used to compress audio.} \\ \\
    \circled{9} & Hybrid Transformation & \makecell[l]{
        We use a combination of methods Gaussian Noise, Time Mask, pitch Shift, and MP3 \\
        Compression to process the audio.} \\ \\
    \circled{10} & Spec Frequency Mask & \makecell[l]{
        Mask a set of frequencies in a spectrogram.} \\ \\
    \circled{11} & Band-Pass Filter & \makecell[l]{
        We use a band-pass filter so that only the sounds in the frequency range from 100Hz \\
        to 6000Hz are retained in the audio.} \\ \\
    \circled{12} & High-Pass Filter & \makecell[l]{
        We apply a high-pass filter to remove the low-frequency portion of the audio.} \\ \\
    \circled{13} & Low-Pass Filter & \makecell[l]{
        We apply a low-pass filter to remove the high-frequency portion of the audio.} \\
    \bottomrule
  \end{tabular}
  }
  \label{tab_trans_describe}
\end{table*}

\end{document}